\documentclass[aps,prb,,amsmath,amssymb,floatfix,twocolumn]{revtex4-1}
\usepackage{times}
\usepackage{bbold}
\usepackage{graphicx}
\usepackage{color}
\usepackage{hyperref}
\usepackage{wasysym}

\begin{document}

\title{Spin dynamics of possible density wave states in the pseudogap phase of the high temperature superconductors}
\author{Chen-Hsuan Hsu}
\affiliation{Department of Physics and Astronomy, University of
California Los Angeles\\ Los Angeles, California 90095-1547}
\author{Zhiqiang Wang}
\affiliation{Department of Physics and Astronomy, University of
California Los Angeles\\ Los Angeles, California 90095-1547}
\author{Sudip Chakravarty}
\affiliation{Department of Physics and Astronomy, University of
California Los Angeles\\ Los Angeles, California 90095-1547}

\date{\today}

\begin{abstract}
In a recent inelastic neutron scattering experiment in the pseudogap state of the high temperature superconductor $\mathrm{YBa_{2}Cu_{3}O_{6.6}}$ an unusual `vertical' dispersion of the spin excitations with a large in-plane anisotropy was observed. In this paper we discuss in detail  the spin susceptibility of the singlet $d$-density wave, the triplet $d$-density wave, as well as the more common spin density wave orders with hopping anisotropies. From  numerical calculations within the framework of random phase approximation, we find nearly vertical dispersion relations for spin excitations with anisotropic incommensurability at low energy $\omega \le 90~meV$, which are reminiscent of the experiments. At very high energy $\omega \ge 165~meV$, we also find energy-dependent incommensurability. Although there are some important difference between the three cases, unpolarized neutron measurements cannot discriminate between these alternate possibilities; the vertical dispersion, however, is a distinct  feature of all three density wave states in contrast to the superconducting state, which shows an hour-glass shape dispersion.

\end{abstract}


\maketitle

\section{Introduction}
The pseudogap state of high temperature superconductors has been studied with numerous experimental tools, yet its origin is not  resolved.~\cite{Norman:2005} One view proposes that the pseudogap state is a  particle-hole condensate, a density wave. Of all  such states that break  translational  symmetry and has strong momentum dependence of the type $d_{x^{2}-y^{2}}$, two candidate density wave orders that can couple to inelastic neutron scattering have been  proposed: the singlet $d_{x^{2}-y^{2}}$-density wave (sDDW),~\cite{Chakravarty:2001} corresponding to angular momentum $\ell=2$ but a spin singlet, and the spin density wave order (SDW); in the general classification of density wave orders,~\cite{Nayak:2000} the latter corresponds to $\ell =0$ but a spin triplet. In addition to the sDDW order, its triplet counterpart~\cite{Nersesyan:1991} (tDDW) $i\sigma{d_{x^{2}-y^{2}}}$, where $\sigma=\pm 1$ corresponding to up and down spins with the $\hat{z}$ axis as the axis of spin quantization   has also interesting properties and deserves more attention.~\cite{Nersesyan:1991} Recently, Fujimoto proposed that a triplet $d$-wave particle-hole condensate may be realized in the hidden order state of the URu$_2$Si$_2$ system~\cite{Fujimoto:2011}. Since high-$T_c$ superconductors have a rich phase diagram, which hosts many possible competing orders, it is both important and interesting to examine the properties of various density wave order parameters of higher angular momentum. In this paper we  discuss the three order parameters mentioned above. In addition, we note that a singlet chiral $i{d_{x^{2}-y^{2}}}+d_{xy}$-density wave~\cite{Tewari:2008}  as well as $i\sigma{d_{x^{2}-y^{2}}}+d_{xy}$ density wave states with interesting topological properties have been explored.~\cite{Hsu:2011} Owing to limitations of space, we do not discuss these order parameters here.

Inelastic neutron scattering can directly  probe magnetic excitations. The scattering cross-section is proportional to the magnetic structure factor, which is proportional to the imaginary part of the dynamic spin susceptibility via the fluctuation-dissipation theorem~\cite{Aeppli:1997}. Thus, a calculation of the spin susceptibility will provide a link between theoretical models and neutron scattering experiments. 

In particular, we want to address a recent experiment in  underdoped YBa$_{2}$Cu$_{3}$O$_{6.6}$. The most striking aspect of this experiment is a vertical dispersion relation of the spin excitations with a large in-plane anisotropy in the pseudogap state in contrast to the `hour glass' dispersion  observed in the superconducting state.~\cite{Hinkov:2007} The qualitatively different behavior between the  superconducting and the pseudogap states suggests different mechanisms.  Motivated by the experimental observations, we study the spin susceptibility of the three density wave orders mentioned above with hopping anisotropy, which breaks $C_{4}$ rotational and mimics an `electron nematic' state.~\cite{Yao:2011} In a phenomenological model, we set the hopping terms to be anisotropic along $a$- and $b$-axes, and study the energy-momentum dispersion relations of the dynamical spin susceptibility.

The structure of this paper is as follows: in Sec. \ref{Sec:sDDW}, we sketch the calculation of the spin susceptibility, and discuss the numerical results of the sDDW order. In Sec. \ref{Sec:tDDW}, we discuss the numerical results of the tDDW order. In Sec. \ref{Sec:SDW}, we also discuss the numerical results of the SDW order. To make the paper succinct and more accessible, the explicit forms of the spin susceptibility are shown in Appendix \ref{Sec:chi}.

\section{\label{Sec:sDDW}Spin Susceptibility: singlet DDW}
% sDDW $id_{x^{2}-y^{2}}$ density wave order
In this section we set up the calculation of the spin susceptibility using sDDW as an example. In the following sections we will give the results of the other order parameters.
To capture the in-plane anisotropic feature of the pseudogap state in the neutron scattering experiment, we consider the sDDW order with anisotropic hopping terms. In the momentum space, the order parameter can be written in terms of the fermion operators as 
\begin{equation}
\langle c_{k+Q,\alpha}^{\dagger} c_{k,\beta}\rangle \propto i \delta_{\alpha\beta} W_{k}
\end{equation}
with $W_k \equiv \frac{W_0}{2} \left[ \cos (k_x a) - \cos (k_y b) \right]$, where $a$ and $b$ are lattice constants. For orthorhombic YBa$_{2}$Cu$_{3}$O$_{6.6}$, $a$ and $b$ are unequal, but the difference is very small. ($a=3.82 \AA, b=3.87 \AA$.) 

The two-dimensional mean field Hamiltonian will be
\begin{eqnarray}
\mathcal{H}_{sDDW}&=&\sum_{\sigma} \sum_{k} \left( \epsilon_k c_{k,\sigma}^{\dagger} c_{k,\sigma} + \epsilon_{k+Q} c_{k+Q,\sigma}^{\dagger} c_{k+Q,\sigma} \right. \nonumber\\
&& \left.  + iW_k c_{k,\sigma}^{\dagger} c_{k+Q,\sigma} + h.c. \right),
\end{eqnarray}
where the summation is over the reduced Brilloin Zone (RBZ) bounded by $(k_y b)\pm (k_x a)= \pm \pi$, $Q=(\pi/a,\pi/b)$ is the nesting vector, and $\epsilon_{k} \equiv \epsilon_{1k}+\epsilon_{2k}$ with~\cite{Pavarini:2001}
\begin{eqnarray}
\epsilon_{1k}&\equiv& -2t \left[ (1+r)\cos (k_x a)+ (1-r) \cos (k_y b) \right],\\
\epsilon_{2k}&\equiv& 4t^{\prime} \cos (k_x a) \cos (k_y b) -\mu \nonumber\\
&& -2t^{\prime\prime}\left[ (1+r)\cos (2k_x a) + (1-r)\cos (2k_y b) \right].
\end{eqnarray}
For $r\neq 0$, we have anisotropic hopping terms which breaks four-fold rotational symmetry. Note that although the anisotropy also modifies the next nearest neighbor hopping, it is simply a parameter and is defined as $t^{\prime}$ in our model. The eigenvalues of the Hamiltonian are
$\lambda_{k,\pm} = \epsilon_{2k} \pm E_{k}$ with $E_{k} \equiv \sqrt{\epsilon_{1k}^2 + W_k^2}$.

The one-loop spin susceptibility in the momentum and Matsubara frequency space is defined as, $N$ being the number of lattice sites,  
\begin{equation}
\chi_0^{ij}(q,q',i \omega_n) = -{\frac {1}{N}} \int^{\beta}_{0} d \tau e^{i \omega_n \tau} \langle T S^{i}_{q}(\tau) S^{j}_{-q'} \rangle,
\end{equation}
where $i,j=x,y,z$, $\tau$ is the imaginary time, $T$ is the time-ordering symbol, and the spin operators are
\begin{equation}
S^{i}_{q} \equiv \sum_{k,\alpha,\beta} c_{k+q,\alpha}^{\dagger} \hat{\mathbf{\sigma}}^{i}_{\alpha\beta} c_{k,\beta}.
\end{equation}
Here $\hat{\mathbf{\sigma}}_{\alpha\beta}$ are the Pauli matrices with spin indices $\alpha$ and $\beta$. We can define the longitudinal and the transverse susceptibilities as $\chi_0^{zz}(q,q',\omega)$ and $\chi_0^{+-}(q,q',\omega)$, respectively, with $S_{q}^{\pm}\equiv S_{q}^{x}\pm iS_{q}^{y}$ and $i \omega_n \rightarrow \omega + i \delta$. For unpolarized measurements, the scattering intensity, $I$, contains both the spin-flip and the non-spin-flip channels, $ I \propto (\chi^{zz} + 2\chi^{+-})/3$. However, in this paper we will present the longitudinal and transverse susceptibilities separately so that it can provide more information about the polarized neutron scattering experiments, which may be  achieved in the future. 

For the sDDW order, $\chi_0^{zz}(q,q',\omega)=2\chi_0^{+-}(q,q',\omega)$ because up-spin and down-spin parts of the Hamiltonian are identical. The explicit form of the one-loop susceptibility is shown in  Eq.~(\ref{Eq-sDDWlong}-~\ref{Eq-sDDW-off}), and we apply random phase approximation (RPA) to obtain the RPA susceptibility as shown in Eq.~(\ref{Eq-sDDWRPA1}-\ref{Eq-sDDWRPA2}) in Appendix \ref{Sec:chi}.
For illustrative purposes, we set $t=0.15~eV$, $t^{\prime}=0.32t$, $t^{\prime\prime}=0.1t^{\prime}$, $W_0=0.65t$, $r=-0.1$, and $k_{B}T=0.05t$. The chemical potential is set to $\mu=-0.805t$ in order to obtain a hole doping level of $n_{h}\approx 10.07\%$, approximately the doping level in the experiment. Other similar choices of the parameters will not change the conclusions.

\begin{figure}[htbp]
\begin{center}
\includegraphics[width=\linewidth]{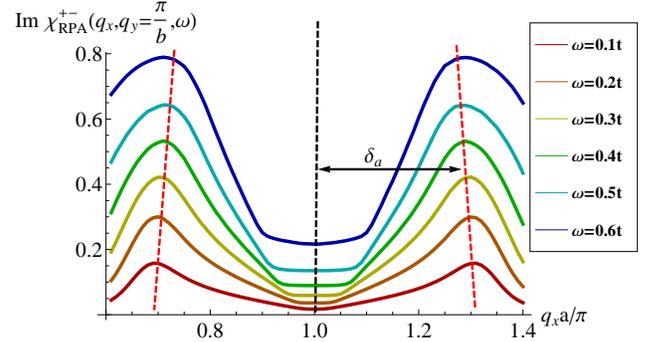}
\caption{(Color online) Constant energy cuts of Im$\chi^{+-}_{RPA}(q,\omega)$ along $a^{*}$-axis when $q_{y}=\pi/b$ and $0.1t \le \omega \le 0.6t$ for the sDDW order. The weakly energy-dependent incommensurate peak positions are marked with red dashed lines. The results of Im$\chi^{zz}_{RPA}(q,\omega)$ are similar and omitted.}
\label{Fig-sDDWx}
\end{center}
\end{figure}

% The numerical results of sDDW order 
In Fig.~\ref{Fig-sDDWx}, the constant energy cuts of the imaginary part of the transverse spin susceptibility along $a^{*}$-axis for $\omega \le 0.6t$ are plotted. The results along $b^{*}$-axis are similar and are not shown here. Away from $Q=(\pi/a,\pi/b)$, the magnetic excitations are peaked at the incommensurate positions $(q_{x} a, q_{y} b) = (\pi \pm \delta_{a},\pi)$ and $(\pi,\pi \pm \delta_{b})$, where we define the incommensurability $\delta_{a}$ and $\delta_{b}$ along $a^{*}$- and $b^{*}$-axes, respectively. 
From the numerical results, one finds that $\delta_{a}$ and $\delta_{b}$ are weakly energy dependent, similar to the inelastic neutron scattering experiment~\cite{Hinkov:2007}. Furthermore, a prominent anisotropy in the incommensurability $\delta_{b}<\delta_{a}$ can be seen. With the hopping anisotropy $r=-0.1$, we obtain $\delta_{a}\approx (0.30\pm 0.01)\pi$ and $\delta_{b}\approx (0.235\pm 0.015)\pi$, which gives $\delta_{b}/\delta_{a} \approx 0.78$, which would be again similar  to $\delta_{b}/\delta_{a} \approx 0.6$ reported  in the neutron scattering experiments.~\cite{Hinkov:2007} 

\begin{figure}[htbp]
\begin{center}
\includegraphics[width=\linewidth]{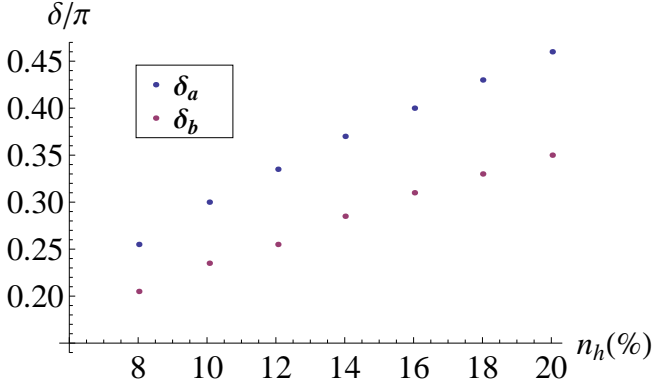}
\caption{Doping-dependence of incommensurability $\delta_{a}$ and $\delta_{b}$. Here $\mu$ is adjusted to obtain different doping levels, and all the other parameters are the same as in Fig.~\ref{Fig-sDDWx}.}
\label{Fig-doping}
\end{center}
\end{figure}

One may further adjust the parameters of this model to fit the experimental data, but that is not the goal of this paper. 
We have varied the chemical potential, $\mu$, to check how the dispersion relations vary with  hole doping;   results for different doping levels are qualitatively similar. In the doping range $8\% \le n_{h} \le 20\%$, there are always weakly energy-dependent incommensurate excitations, and the incommensurability $\delta_{a}$ and $\delta_{b}$ increase with increasing doping level $n_{h}$ as shown in Fig.~\ref{Fig-doping}.

\begin{figure}[htbp]
\begin{center}
\includegraphics[width=\linewidth]{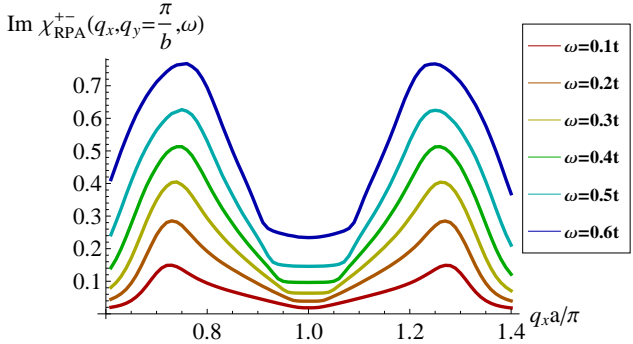}
\caption{(Color online) (Color online) Constant energy cuts of Im$\chi^{+-}_{RPA}(q,\omega)$ along $a^{*}$-axis when $q_{y}=\pi/b$ for the sDDW order. Here $r=0$, $\mu=-0.806t$, and all the other parameters are the same as in Fig.~\ref{Fig-sDDWx}.}
\label{Fig-sDDWiso}
\end{center}
\end{figure}

Note that hopping anisotropy is not necessary for the existence of the nearly vertical dispersions. To demonstrate this, the numerical results with isotropic hopping are plotted in Fig.~\ref{Fig-sDDWiso}. Here $r$ is set to $0$, $\mu=-0.806t$, and the hole doping level is $n_{h}=10.03\%$. All the other parameters are the same as in Fig.~\ref{Fig-sDDWx}. One can still find nearly vertical dispersions with incommensurability $\delta_{a}\approx (0.255\pm 0.015)\pi$ even without the hopping anisotropy.~\cite{Jiang:2011}
 
\begin{figure}[htbp]
\begin{center}
\includegraphics[width=\linewidth]{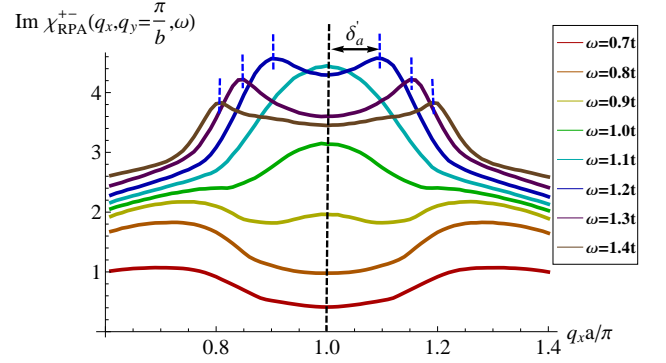}
\caption{ (color online) Constant energy cuts of Im$\chi^{+-}_{RPA}(q,\omega)$ along $a^{*}$-axis when $q_{y}=\pi/b$ and $0.7t \le \omega \le 1.4t$ for the sDDW order. The energy-dependent incommensurate peak positions are marked with blue dashed lines.}
\label{Fig-sDDWx-hi}
\end{center}
\end{figure}

The neutron scattering experiments show vertical dispersions in the energy range $30~meV \le \omega \le 60~meV$,~\cite{Hinkov:2007} and the numerical results exhibit a nearly vertical dispersions up to $\omega \le 0.6t = 90~meV$ with the chosen parameters, which are similar to experiments. It is interesting to see how the excitation peaks evolve at higher energies, so in Fig.~\ref{Fig-sDDWx-hi} we present the numerical results along the $a^{*}$-axis for $0.7t \le \omega \le 1.4t$, where all the parameters are the same as in Fig.~\ref{Fig-sDDWx} except for the energy $\omega$. The results along $b^{*}$-axis are again so similar that they are not shown here. In Fig.~\ref{Fig-sDDWx-hi}, one finds that the high energy spin excitations are strongly energy dependent. The incommensurate peaks move toward $q=Q$ in the range  $0.7t \le \omega \le 0.9t$, and eventually disappear at $\omega \approx 1.0t$, where the intensity around $q=Q$ is enhanced. When $\omega \approx 1.1t$, a central peak emerges at the commensurate position $q=Q$. As the energy is further increased, the central peak splits into to two peaks deviating  from $Q$ with incommensurability $\delta_{a}^{\prime}$ and $\delta_{b}^{\prime}$, which are marked by dashed lines. Unlike the low-energy incommensurability $\delta_{a}$ and $\delta_{b}$, $\delta_{a}^{\prime}$ and $\delta_{b}^{\prime}$ are energy-dependent and increase with increasing energy. Note that to observe $\delta_{a}^{\prime}$ and $\delta_{b}^{\prime}$, the neutron scattering experiment needs to be performed with very high energy ($\omega \ge 1.1t = 165~meV$).

\begin{figure}[htbp]
\begin{center}
\includegraphics[width=\linewidth]{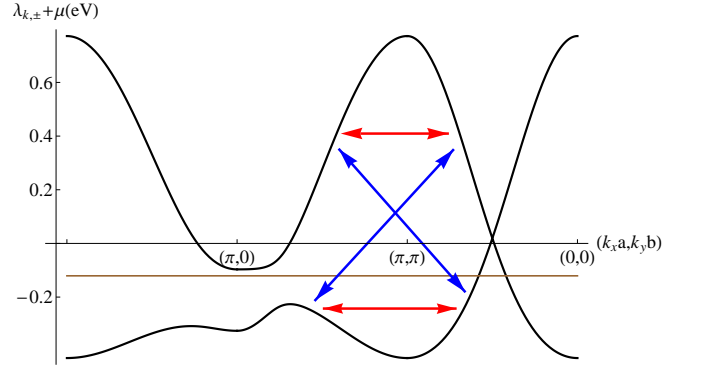}
\caption{(Color online)Energy spectrum $(\lambda_{k,\pm}+\mu)$ of the sDDW system as $(k_{x}a,k_{y}b)$ goes along the route: $(0,0) \rightarrow (\pi,0) \rightarrow (\pi,\pi) \rightarrow (0,0)$. The blue (red) arrows indicate the interband (intraband) scattering, and the brown line is the chemical potential $\mu$. The parameters are the same as in Fig.~ \ref{Fig-sDDWx}.}
\label{Fig-Scatt}
\end{center}
\end{figure}

The reason for the unusual vertical dispersions at low energies and a different behavior at high energies can be understood by examining the imaginary part of Eq.~(\ref{Eq-sDDW-diag}). In this equation, the first two terms are interband contribution arising from the scattering from the upper band ($\epsilon_{2k}+E_{k}$) to the lower band ($\epsilon_{2k+q}-E_{k+q}$), and the scattering from the lower band ($\epsilon_{2k}-E_{k}$) to the upper band ($\epsilon_{2k+q}+E_{k+q}$). The last two terms, on the other hand, are intraband scattering. For the purpose of illustration, an example of the band structure and the scattering process is plotted in Fig.~~\ref{Fig-Scatt}, where the interband and intraband scattering are shown with arrows.

\begin{figure}[htbp]
\begin{center}
\includegraphics[width=\linewidth]{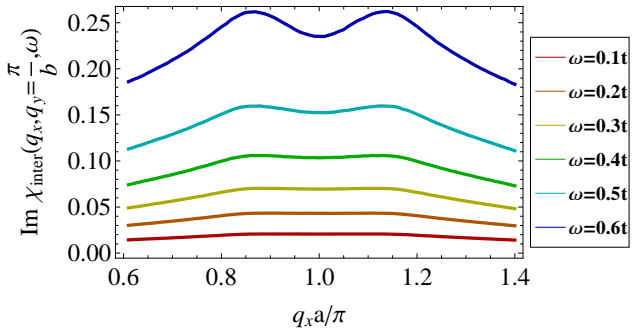}
\caption{ (Color online) Constant energy cuts of the interband terms of Im$\chi_{diag}(q,\omega)$ in Eq.~(\ref{Eq-sDDW-diag}) along $a^{*}$-axis when $q_{y}=\pi/b$ for $0.1t \le \omega \le 0.6t$.}
\label{Fig-interband}
\end{center}
\end{figure}

\begin{figure}[htbp]
\begin{center}
\includegraphics[width=\linewidth]{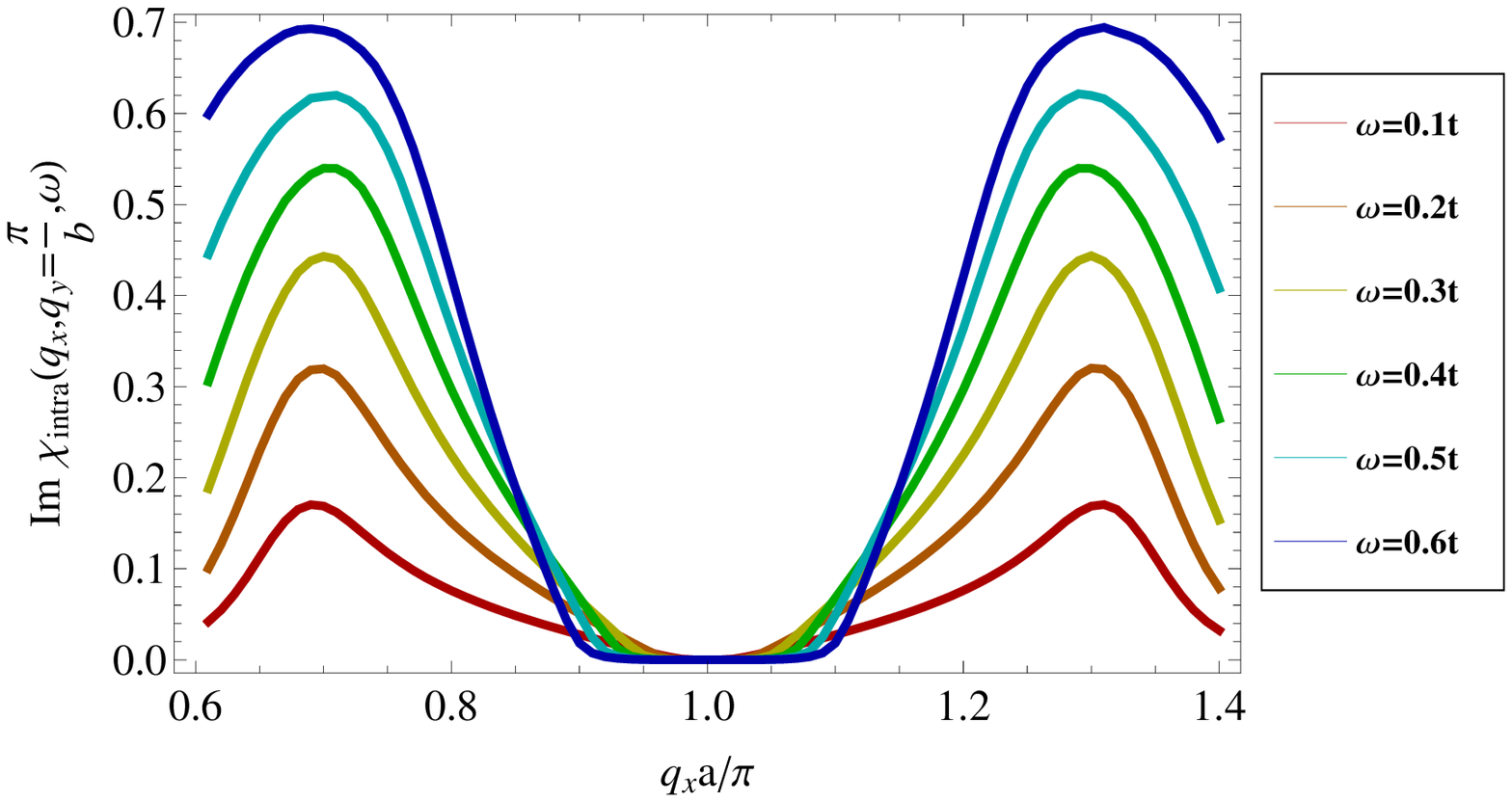}
\caption{(Color online) Constant energy cuts of the intraband terms of Im$\chi_{diag}(q,\omega)$ in Eq.~(\ref{Eq-sDDW-diag}) along $a^{*}$-axis when $q_{y}=\pi/b$ for $0.1t \le \omega \le 0.6t$.}
\label{Fig-intraband}
\end{center}
\end{figure}

The interband and intraband terms of Eq.~(\ref{Eq-sDDW-diag}) for $0.1t\le\omega\le 0.6t$ are plotted in Fig.~\ref{Fig-interband} and Fig.~\ref{Fig-intraband}, respectively. The results for higher energy $0.7t\le\omega\le 1.4t$  are not shown because they are very similar. From Fig.~\ref{Fig-interband} and Fig.~\ref{Fig-intraband}, one finds that the intensity near $q=Q$ is mainly from the contribution of the interband terms, whereas the contribution of the intraband terms arise when $q$ is away from $Q$.  From Eq.~(\ref{Eq-sDDW-diag}), we can see that at $q=Q$, the intraband terms vanish and only the interband terms contribute, leading to magnetic excitations peaked around $\omega \approx 1.1t$. In the vicinity of $q=Q$, interband terms still dominate, and we may expand them to  first order  in $\delta q\equiv |q-Q|$ and obtain  
\begin{eqnarray*}
 &\frac{-\pi}{N} \sum_{k} \left[n_F(\epsilon_{2k} \pm E_k) - n_F(\epsilon_{2k+q} \mp E_{k+q})\right] \times \\
&\delta(\omega - \epsilon_{2k} \mp E_k + \epsilon_{2k+q} \mp E_{k+q})\\
\simeq& \frac{\pi}{N} \sum_{k} \left[ n_F(\epsilon_{2k} \mp E_k) - n_F(\epsilon_{2k} \pm E_{k})\right. \\
& \left. + \frac{\partial n_F(E)}{\partial E}|_{E=\epsilon_{2k} \mp E_k} \vec{\triangledown}_k (\epsilon_{2k} \mp E_k)\cdot \delta q   \right]
\times \\
&\delta(\omega \mp 2E_k + \vec{\triangledown}_k (\epsilon_{2k} \mp E_k)\cdot \delta q ),
\end{eqnarray*}
which will be peaked at $\delta q = (\pm 2E_{k}-\omega)/\left[ \vec{\triangledown}_{k} (\epsilon_{2k} \mp E_k) \right]$. However, for low energies, the energy conservation condition cannot be satisfied unless $E_k$ is very small, which diminishes the difference between the Fermi functions and thus suppresses the intensity. Therefore, there is no enhanced peak in the vicinity of $q=Q$ for low energies. For higher energies, the energy conservation factor will be satisfied, and the intensity at the incommesurate positions ($\delta_{a}^{\prime}$ and $\delta_{b}^{\prime}$) will be enhanced and the excitation peaks can be seen as $\omega \apprge 1.1t$ in Fig.~\ref{Fig-sDDWx-hi}.

In contrast, away from $q=Q$, the intraband terms dominate. The peak positions of the energy conservation factor, $\delta(\omega - \epsilon_{2k} \mp E_k + \epsilon_{2k+q} \pm E_{k+q})$, move away from $Q$ with increasing $\omega$. On the other hand, the coherence factor $\left[ 1+  (\epsilon_{1k} \epsilon_{1k+q} + W_k W_{k+q}) / (E_k E_{k+q}) \right]$ vanishes at $q=Q$ and develops with increasing $|q-Q|$. For the chosen parameters, the energy dependence of these two opposite effects almost cancels out in the energy range $0 \le \omega \le 0.6t$, leading to the weakly energy-dependent positions of local maxima ($\delta_{a}$ and $\delta_{b}$) as in Fig.~\ref{Fig-intraband}. Such a dispersionless feature is  sensitive to the parameters because it depends on whether the contribution of the intraband terms overcomes that of the interband terms away from $Q$. The nature of the excitation peaks due to the interband terms is distinct from the intraband terms. The dominant contribution of the interband terms are determined by the energy conservation factor and the Fermi functions, leading to sharper excitation peaks at $(\pi \pm \delta_{a}^{\prime},\pi)$ and $(\pi,\pi \pm \delta_{b}^{\prime})$, whereas the intraband terms also depend on the coherence factor, resulting in relatively broadened local maxima instead of sharp peaks at $(\pi \pm \delta_{a},\pi)$ and $(\pi,\pi \pm \delta_{b})$.

\section{\label{Sec:tDDW}The triplet $d$-density wave order}
% tDDW $i\sigma d_{x^2-y^2}$ density wave order
 
We now consider the tDDW order, and choose the spin quantization axis to be the $z$-axis without any loss of generality, that is,
\begin{equation}
\langle c_{k+Q,\alpha}^{\dagger} c_{k,\beta}\rangle \propto i (\hat{d}\cdot \vec{\sigma}_{\alpha\beta})W_{k}=i (\hat{z}\cdot \vec{\sigma}_{\alpha\beta})W_{k}.
\end{equation}
The tDDW mean field Hamiltonian is therefore
\begin{eqnarray}
\mathcal{H}_{tDDW}&=&\sum_{\sigma} \sum_{k} \left( \epsilon_k c_{k,\sigma}^{\dagger} c_{k,\sigma} + \epsilon_{k+Q} c_{k+Q,\sigma}^{\dagger} c_{k+Q,\sigma} \right. \nonumber\\
&& \left.  + i\sigma W_k c_{k,\sigma}^{\dagger} c_{k+Q,\sigma} + h.c. \right),
\end{eqnarray}
which has the same eigenvalues as the sDDW Hamiltonian. The explicit form of the one-loop and RPA susceptibilities are given in Eq.~(\ref{Eq-tDDWlong}-~\ref{Eq-tDDWRPA2}) in Appendix \ref{Sec:chi}.

\begin{figure}[htbp]
\begin{center}
\includegraphics[width=\linewidth]{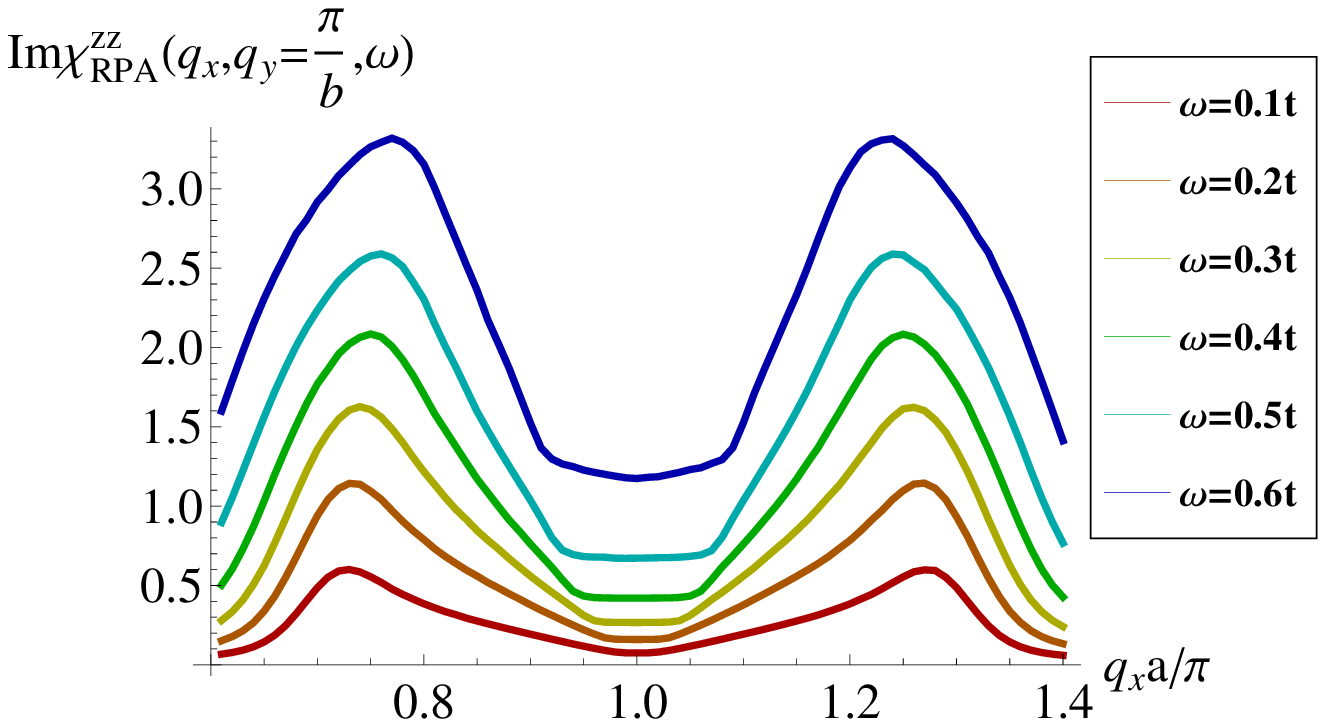}
\includegraphics[width=\linewidth]{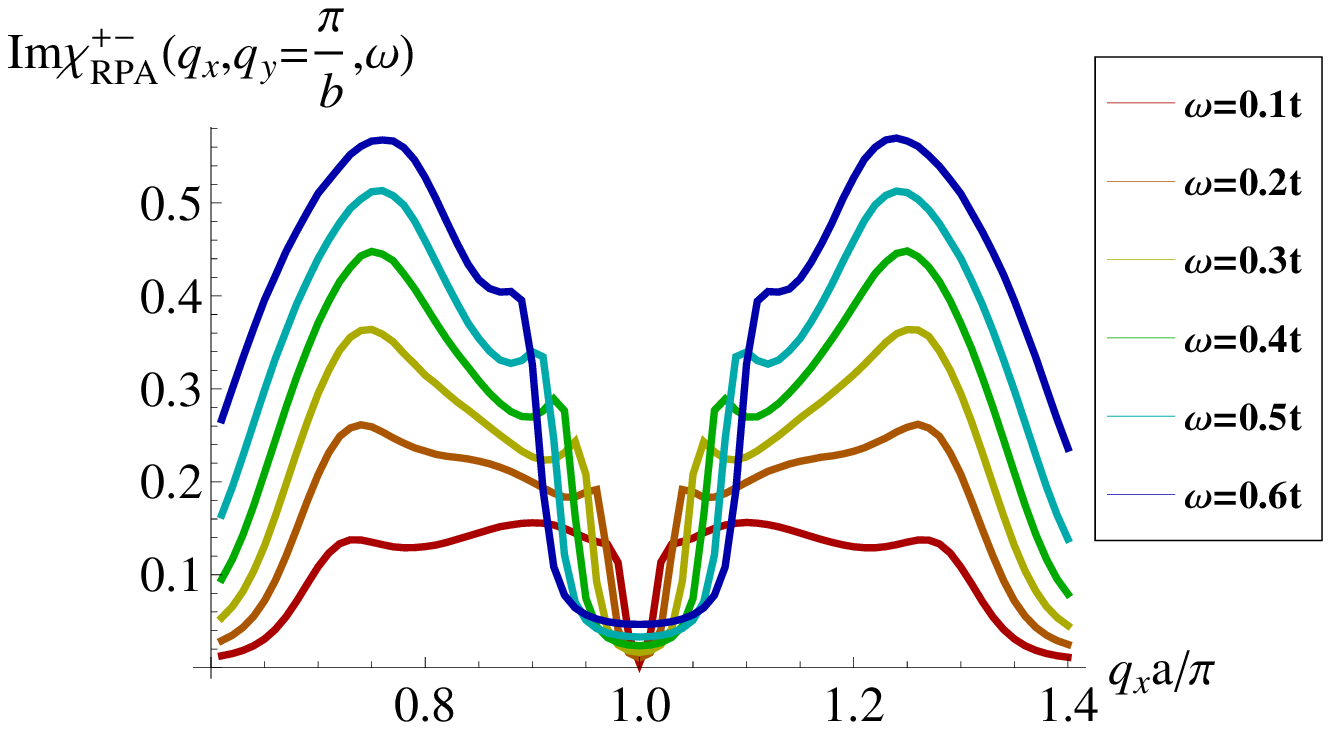}
\caption{(Color online) Constant energy cuts of Im$\chi_{RPA}^{zz}(q,\omega)$ (upper) and Im$\chi_{RPA}^{+-}(q,\omega)$ (lower) for the tDDW order along $a^{*}$-axis when $q_y=\pi/b$. The parameters are the same as in Fig.~\ref{Fig-sDDWiso}.}
\label{Fig-tDDW}
\end{center}
\end{figure}

The constant energy cuts of the imaginary part of the spin susceptibility of the tDDW order along $a^{*}$-axis are shown in Fig.~\ref{Fig-tDDW}. The hopping anisotropy $r$ is set to $0$ for simplicity and the parameters are the same as in Fig.~\ref{Fig-sDDWiso}. The longitudinal susceptibility behaves similar to the sDDW order whereas the transverse susceptibility is significantly different  in the vicinity of $q=Q$. In comparison with the sDDW order, the intensity of Im$\chi^{+-}_{RPA}(q,\omega)$ of the tDDW order is suppressed in the vicinity of $q=Q$. The intensity exhibits a V-shaped curve around $q=Q$ at $\omega=0.1t$, which evolves gradually to a U-shaped curve at $\omega=0.6t$. Here we can also see the nearly vertical dispersion of the incommensurate spin excitations $\delta_a \approx (0.255\pm 0.015)\pi$. Notice that for unpolarized measurements, with $I \propto (\chi^{zz} + 2\chi^{+-})/3$,  there will still be the vertical dispersion away from $q=Q$.

The difference between the sDDW and tDDW order is that in $\chi_{\text{diag}}^{+-}(q,\omega)$ of the tDDW order, Eq.~(\ref{Eq-tDDW-T-diag}), the $W_{k}W_{k+q}$ term of the coherence factor changes sign and reduces the interband contribution. As a result, the intensity in the vicinity of $q=Q$ is suppressed. The significant difference between the transverse and the longitudinal susceptibilities should permit one to distinguish the singlet and the triplet orders in spin-polarized measurements. On the other hand, the sign change of $W_{k}W_{k+q}$ does not affect the intraband terms as much as the interband terms, so the nearly vertical dispersions due to the intraband contribution can still be seen away from $q=Q$.

\section{\label{Sec:SDW}The spin density wave order}

Finally, we also consider the SDW order, which has the order parameter 
\begin{equation}
\langle c_{k+Q,\alpha}^{\dagger} c_{k,\beta}\rangle \propto (\hat{z}\cdot \vec{\sigma}_{\alpha\beta}) \Delta_{s}.
\end{equation}
The SDW mean field Hamiltonian will be
\begin{eqnarray}
\mathcal{H}_{SDW}&=&\sum_{\sigma} \sum_{k} \left( \epsilon_k c_{k,\sigma}^{\dagger} c_{k,\sigma} + \epsilon_{k+Q} c_{k+Q,\sigma}^{\dagger} c_{k+Q,\sigma} \right. \nonumber\\
&& \left.  + \sigma \Delta_{s} c_{k,\sigma}^{\dagger} c_{k+Q,\sigma} + h.c. \right),
\end{eqnarray}
where the eigenvalues now become $\lambda^{S}_{k,\pm}=\epsilon_{2k}\pm E^{S}_{k}$ with $E^{S}_{k} \equiv \sqrt{\epsilon_{1k}^2 + \Delta_{s}^2}$. The explicit forms of the spin susceptibilities are given in Eq.~(\ref{Eq-SDWlong}-~\ref{Eq-SDW-T-off}) in Appendix \ref{Sec:chi}.

\begin{figure}[htbp]
\centering
\includegraphics[width=\linewidth]{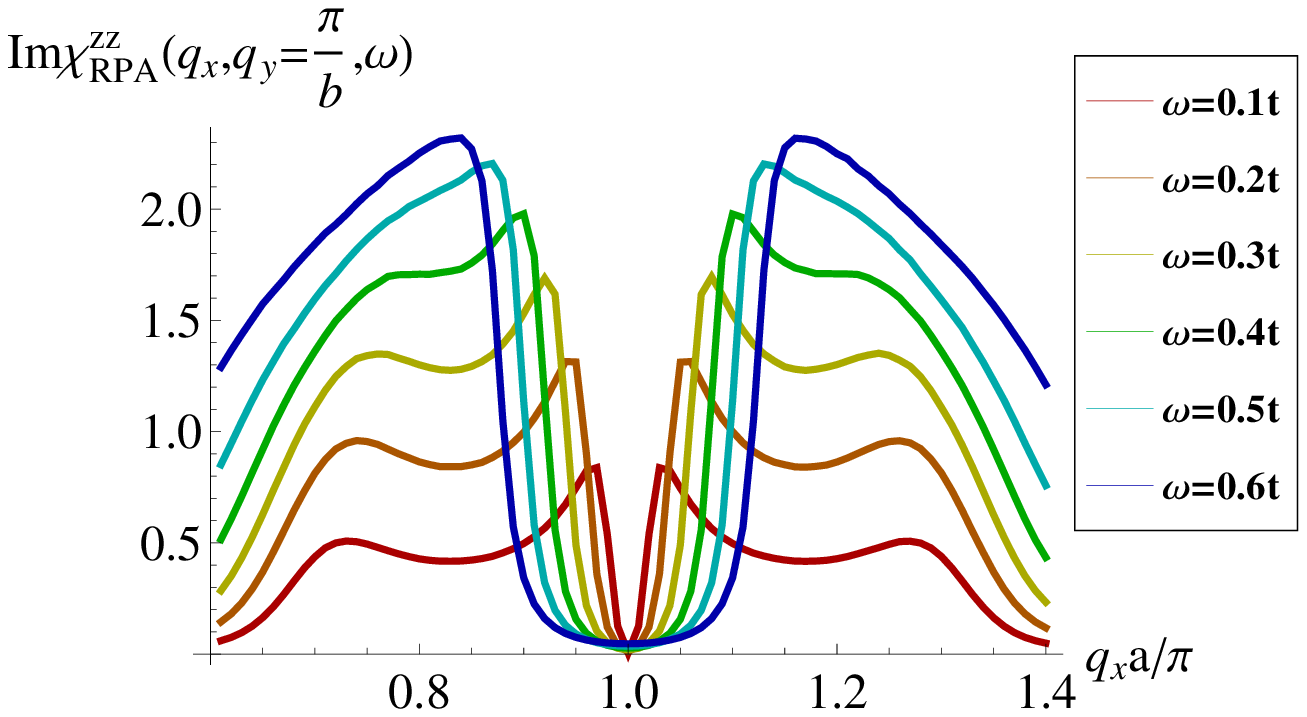}
\includegraphics[width=\linewidth]{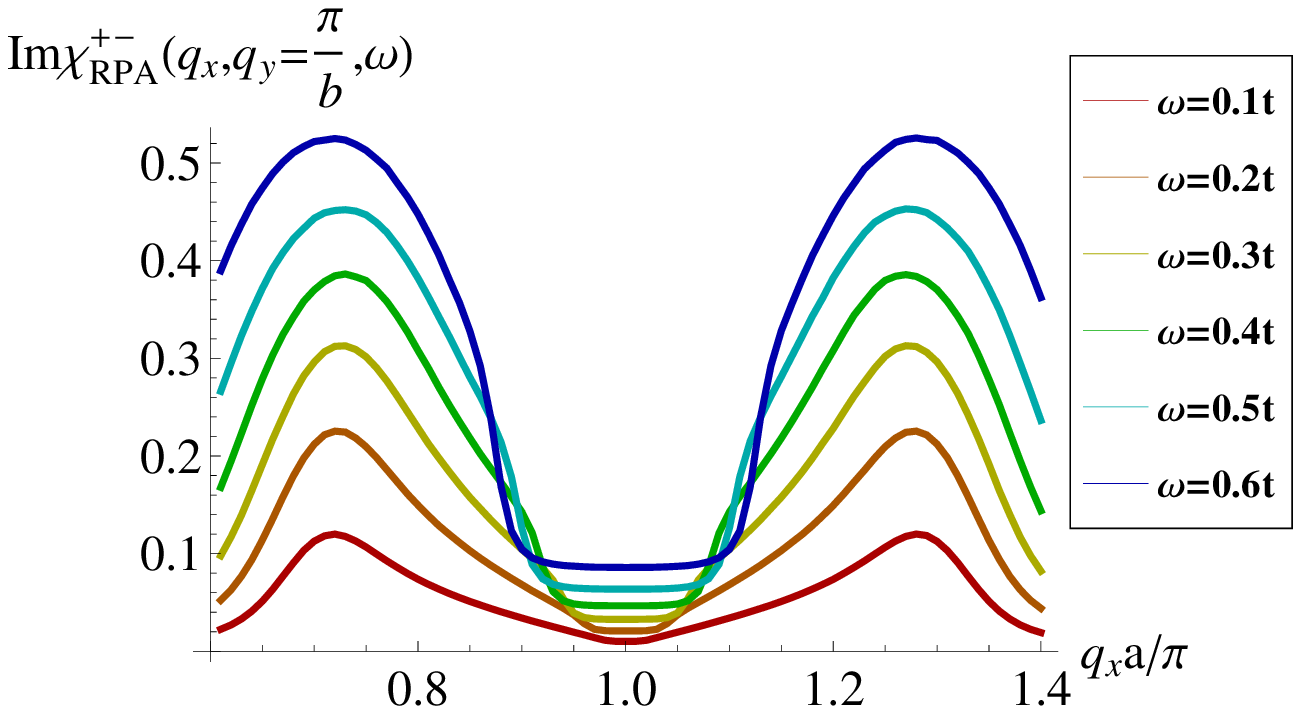}
\caption{Constant energy cuts of Im$\chi_{RPA}^{zz}(q,\omega)$ (upper) and Im$\chi_{RPA}^{+-}(q,\omega)$ (lower) for the SDW order along $a^{*}$-axis when $q_y=\pi/b$. Here $\Delta_{s}=0.65t$, $\mu=-1.026t$, and the other parameters are the same as in Fig.~\ref{Fig-tDDW}.}
\label{Fig-SDW}
\end{figure}

The constant energy cuts of Im$\chi_{RPA}^{zz}(q,\omega)$ and Im$\chi_{RPA}^{+-}(q,\omega)$ for the SDW order along $a^{*}$-axis are plotted in Fig.~\ref{Fig-SDW}. Here we set the SDW gap to be $\Delta_{s}=0.65t$ and $\mu=-1.026t$. The hole doping level is $n_{h}=10.02\%$. The results are interesting: 
Im$\chi_{RPA}^{zz}(q,\omega)$ and Im$\chi_{RPA}^{+-}(q,\omega)$ for  SDW order seem to be `interchanged' in comparison with those for the tDDW order in Fig.~\ref{Fig-tDDW}. In addition to  this interchange, there is also a  difference in the intensity around $q=Q$ between  tDDW and SDW,
which could be observed if spin-polarized experiments with high resolution could be achieved, although one cannot be sure because of the non universal nature of this difference. Away from $q=Q$, we can also see the vertical dispersions of the incommensurate spin excitations with $\delta_a \approx 0.28\pi$. Again, for unpolarized measurements, there will still be the vertical dispersion away from $q=Q$.

To understand the swap of the susceptibilities between tDDW and SDW, we should compare Eq.~(\ref{Eq-sDDW-diag}) and Eq.~(\ref{Eq-tDDW-T-diag}) for the tDDW  with Eq.~(\ref{Eq-SDW-L-diag}) and Eq.~(\ref{Eq-SDW-T-diag}) for the SDW;   we can see that at $q=Q$, $W_{k}W_{k+q}=-W^2_{k}$ in  tDDW, and this  leads to a minus sign, while $\Delta^2_{s}$ in  SDW  does not. Therefore, the form of the  coherence factors of  SDW  is opposite to tDDW    in the vicinity of $q=Q$. As a result, the intensity of Im$\chi_{RPA}^{+-}(q,\omega)$ for  SDW  in the vicinity of $q=Q$ is enhanced due to the dominant interband contribution, whereas the intensity of Im$\chi_{RPA}^{zz}(q,\omega)$ is suppressed in the vicinity of $q=Q$. Thus, the  difference in coherence factors leads to the ``interchanging" behavior between  tDDW and SDW;   the different momentum dependence of the order parameters also leads to distinct momentum dependence around $q=Q$. Away from $q=Q$, on the other hand, both Im$\chi_{RPA}^{+-}(q,\omega)$ and Im$\chi_{RPA}^{zz}(q,\omega)$ show vertical dispersion relations due to  intraband contributions.

\section{Conclusion}
In conclusion, we have attempted to provide an explanation of a recent neutron scattering measurement in an underdoped  high temperature superconductor, which point to the fact that the pseudogap state is not a continuation of the superconducting state below $T_{c}$. The salient feature is a vertical dispersion seen above $T_{c}$, in the spin excitations as opposed to an hourglass shape dispersion seen below $T_{c}$.

Although couched in the language Hartree-Fock theory augmented  by RPA,  a thorough analysis of the properties of various alternate order parameters should be a useful guide. We  also checked a band structure to contain electron pockets as well, but the robust aspects of the conclusions were unchanged. The vertical dispersion feature appears to persist in the doping range $8\% \le n_{h} \le 20\%$. At higher energies, we find energy dependent incommensurability due to the interband contributions. We also contrast  the spin dynamics  of the tDDW and SDW orders, which exhibit different features around $q=Q$, which could in principle allow one to identify the spin nature of the underlying phase in  a spin-polarized neutron scattering experiment with high resolution. The  transverse and the longitudinal spin dynamics are interchanged between SDW and tDDW.  In principle, a whole class of higher angular momentum particle-hole condensates are possible. Experimental evidence of these order parameters should be a major step forward.  
The tDDW is such an unconventional hidden order that its discovery would be of great importance. Note that tDDW is even invariant under time reversal. 
 
\begin{acknowledgments}
This work is supported by NSF under Grant No. DMR-1004520. 
\end{acknowledgments}

\appendix
\section{\label{Sec:chi}The explicit forms of spin susceptibility}

In the density wave systems we considered above, the Green's functions form  matrices
\begin{eqnarray*}
{\hat G}_{\sigma}(k, i \omega_m) \equiv \left(
  \begin{array}{cc}
  g_{\sigma}(k, k, i \omega_m)  & g_{\sigma}(k, k+Q, i \omega_m)          \\
  g_{\sigma}(k+Q, k, i \omega_m)& g_{\sigma}(k+Q, k+Q, i \omega_m)        \\
  \end{array}
\right),
\end{eqnarray*}
where 
\begin{eqnarray*}
g_{\sigma}(k, k', i \omega_m) = -\int^{\beta}_{0} d \tau e^{i \omega_m \tau} \langle T c_{k,\sigma} (\tau) c_{k',\sigma}^{\dagger} \rangle.
\end{eqnarray*}

The one-loop spin susceptibility also has diagonal and off-diagonal terms
\begin{align}
\chi_0^{zz}(q,q',i \omega_m) &= \delta_{q,q'} \chi_{\text{diag}}^{zz}(q,i \omega_m) + \delta_{q,q'+Q} \chi_{\text{off}}^{zz}(q,i \omega_m), \nonumber 
\\
\chi_0^{+-}(q,q',i \omega_m) &= \delta_{q,q'} \chi_{\text{diag}}^{+-}(q,i \omega_m) + \delta_{q,q'+Q} \chi_{\text{off}}^{+-}(q,i \omega_m), \nonumber 
\end{align}
where the subscripts `diag' and `off' refer to the diagonal and off-diagonal terms of the one-loop spin susceptibility, respectively. 

With a quadratic Hamiltonian, these terms can be written in terms of the Green's function matrices by applying Wick's theorem, and we have
\small
\begin{eqnarray*}
\chi_{\text{diag}}^{zz}(q,i \omega_m) &=&  {\frac {1}{\beta N}} \sum_{k,n,\sigma} Tr[{\hat G}_{\sigma}(k+q, i \epsilon_n + i \omega_m) {\hat G}_{\sigma}(k, i \epsilon_n)],\\
\chi_{\text{off}}^{zz}(q,i \omega_m) &=&  {\frac {1}{\beta N}} \sum_{k,n,\sigma} \sum_{j\neq l}[{\hat G}_{\sigma}(k+q, i \epsilon_n + i \omega_m) {\hat G}_{\sigma}(k, i \epsilon_n)]_{jl},\\
\chi_{\text{diag}}^{+-}(q,i \omega_m) &=&  {\frac {1}{\beta N}} \sum_{k,n} Tr[{\hat G}_{\uparrow}(k+q, i \epsilon_n + i \omega_m) {\hat G}_{\downarrow}(k, i \epsilon_n)],\\
\chi_{\text{off}}^{+-}(q,i \omega_m) &=&  {\frac {1}{\beta N}} \sum_{k,n} \sum_{j\neq l}[{\hat G}_{\uparrow}(k+q, i \epsilon_n + i \omega_m) {\hat G}_{\downarrow}(k, i \epsilon_n)]_{jl},
\end{eqnarray*}
\normalsize
where $Tr$ is the trace, and ${\hat G}_{\sigma}(k,i\epsilon_n)$ can be obtained from the Hamiltonian.

For sDDW, the up-spin and down-spin components are identical. For $\sigma=\uparrow$ or $\downarrow$, we have
\begin{eqnarray*}
 &&{\hat G}_{\sigma}(k, i \epsilon) =\nonumber \\
& &{\frac {1}{(i \epsilon-\epsilon_{2k})^2-E_k^2}}\left(
  \begin{array}{cc}
   i \epsilon +\epsilon_{1k} -\epsilon_{2k}  &iW_k          \\
   -iW_k     & i \epsilon - \epsilon_{1k} -\epsilon_{2k}         \\
  \end{array}
\right).
\end{eqnarray*}

Therefore, we have
\begin{align}
\chi_0^{zz}(q,q',\omega) &= \delta_{q,q'} \chi_{\text{diag}}(q,\omega) + \delta_{q,q'+Q} \chi_{\text{off}}(q,\omega), \label{Eq-sDDWlong}
\\
\chi_0^{+-}(q,q',\omega) &= \frac{1}{2} \chi_0^{zz}(q,q',\omega), \label{Eq-sDDWtrans}
\end{align}
where
\begin{widetext}
\begin{eqnarray}
\chi_{\text{diag}}(q,\omega)
=&{\frac {1}{N}} \sum_{k} (1- {\frac{\epsilon_{1k} \epsilon_{1k+q} + W_k W_{k+q}}{E_k E_{k+q}}}) \left[{\frac{n_F(\epsilon_{2k} + E_k) - n_F(\epsilon_{2k+q} - E_{k+q})}{\omega - \epsilon_{2k} - E_k + \epsilon_{2k+q} - E_{k+q} + i \delta}} 
+ {\frac{n_F(\epsilon_{2k} - E_k) - n_F(\epsilon_{2k+q} + E_{k+q})}{\omega - \epsilon_{2k} + E_k + \epsilon_{2k+q} + E_{k+q} + i \delta}} \right] \nonumber \\
&+{\frac {1}{N}} \sum_k (1+ {\frac{\epsilon_{1k} \epsilon_{1k+q} + W_k W_{k+q}}{E_k E_{k+q}}}) \left[{\frac{n_F(\epsilon_{2k} + E_k) - n_F(\epsilon_{2k+q} + E_{k+q})}{\omega - \epsilon_{2k} - E_k + \epsilon_{2k+q} + E_{k+q} + i \delta}} 
+ {\frac{n_F(\epsilon_{2k} - E_k) - n_F(\epsilon_{2k+q} - E_{k+q})}{\omega - \epsilon_{2k} + E_k + \epsilon_{2k+q} - E_{k+q} + i \delta}} \right],
\label{Eq-sDDW-diag} 
\end{eqnarray}
\begin{eqnarray}
\chi_{\text{off}}(q,\omega)
=&{\frac {i}{N}} \sum_k (\frac{-\epsilon_{1k} W_{k+q} + \epsilon_{1k+q} W_k }{E_k E_{k+q}}) 
\left[ -\frac{n_F(\epsilon_{2k} + E_k) - n_F(\epsilon_{2k+q} - E_{k+q})}{\omega - \epsilon_{2k} - E_k + \epsilon_{2k+q} - E_{k+q} + i \delta} 
- \frac{n_F(\epsilon_{2k} - E_k) - n_F(\epsilon_{2k+q} + E_{k+q})}{\omega - \epsilon_{2k} + E_k + \epsilon_{2k+q} + E_{k+q} + i \delta} \right. \nonumber \\
& \left. + \frac{n_F(\epsilon_{2k} + E_k) - n_F(\epsilon_{2k+q} + E_{k+q})}{\omega - \epsilon_{2k} - E_k + \epsilon_{2k+q} + E_{k+q} + i \delta}
+ \frac{n_F(\epsilon_{2k} - E_k) - n_F(\epsilon_{2k+q} - E_{k+q})}{\omega - \epsilon_{2k} + E_k + \epsilon_{2k+q} - E_{k+q} + i \delta} \right],
\label{Eq-sDDW-off} 
\end{eqnarray}
\end{widetext}
where $n_F(E)$ is Fermi-Dirac distribution function, and $\delta$ is set to $0.06t$ for the numerical calculation in order to obtain smooth curves. %(also, to incorporate the energy resolution limit in experiments and the broadening due to impurity.)
Applying random phase approximation, we obtain the RPA susceptibility~\cite{Schrieffer:1989}
\begin{align}
\hat{\chi}^{zz}_{RPA}(q,q',\omega) &= \sum_{q_1} \frac{\hat{\chi}^{zz}_0(q,q_1,\omega) }
{\hat{I} - U \hat{\chi}^{zz}_0(q_1,q',\omega)}\label{Eq-sDDWRPA1} \\
\hat{\chi}^{+-}_{RPA}(q,q',\omega) &= \sum_{q_1} \frac{\hat{\chi}^{+-}_0(q,q_1,\omega)} 
{\hat{I} - U \hat{\chi}^{+-}_0(q_1,q',\omega)}
\label{Eq-sDDWRPA2}
\end{align}
where $\hat{\chi}^{zz}_{0}(q,q',\omega)$ and $\hat{\chi}^{+-}_{0}(q,q',\omega)$ are the two by two matrices from Eq.~(\ref{Eq-sDDWlong}) and Eq.~(\ref{Eq-sDDWtrans}), respectively. For the numerical calculation, we set $U=W_0=0.65t$ and compute the imaginary part of the diagonal terms of the RPA susceptibility ($q=q'$). 
%the off-diagonal terms $\hat{\chi}_{RPA}(q,q+Q,\omega)$ are very small.

% tDDW
For the tDDW order, the Green's function matrices become
\begin{eqnarray*}
 &&{\hat G}_{\sigma}(k, i \epsilon) =\nonumber \\
& &{\frac {1}{(i \epsilon-\epsilon_{2k})^2-E_k^2}}\left(
  \begin{array}{cc}
   i \epsilon +\epsilon_{1k} -\epsilon_{2k}  &i\sigma W_k          \\
   -i \sigma W_k     & i \epsilon - \epsilon_{1k} -\epsilon_{2k}         \\
  \end{array}
\right),
\end{eqnarray*}
where $\sigma=+1$ for up-spin and $\sigma=-1$ for down-spin, and the spin susceptibility will become
\begin{align}
\chi_0^{zz}(q,q',\omega) &= \delta_{q,q'} \chi_{\text{diag}}^{zz}(q,\omega), \label{Eq-tDDWlong}\\
\chi_0^{+-}(q,q',\omega) &= \delta_{q,q'} \chi_{\text{diag}}^{+-}(q,\omega) + \delta_{q,q'+Q} \chi_{\text{off}}^{+-}(q,\omega) \label{Eq-tDDWtrans},
\end{align}
where $\chi_{\text{diag}}^{zz}(q,\omega)$ is the same as $\chi_{\text{diag}}(q,\omega)$ in Eq.~(\ref{Eq-sDDW-diag}), and
\begin{widetext}
\begin{eqnarray}
\chi_{\text{diag}}^{+-}(q,\omega)
=&{\frac {1}{2N}} \sum_k (1- {\frac{\epsilon_{1k} \epsilon_{1k+q} - W_k W_{k+q} }{E_k E_{k+q}}}) \left[{\frac{n_F(\epsilon_{2k} + E_k) - n_F(\epsilon_{2k+q} - E_{k+q})}{\omega - \epsilon_{2k} - E_k + \epsilon_{2k+q} - E_{k+q} + i \delta}} 
+ {\frac{n_F(\epsilon_{2k} - E_k) - n_F(\epsilon_{2k+q} + E_{k+q})}{\omega - \epsilon_{2k} + E_k + \epsilon_{2k+q} + E_{k+q} + i \delta}} \right] \nonumber\\
&+{\frac {1}{2N}} \sum_k (1+ {\frac{\epsilon_{1k} \epsilon_{1k+q} - W_k W_{k+q} }{E_k E_{k+q}}})  \left[{\frac{n_F(\epsilon_{2k} + E_k) - n_F(\epsilon_{2k+q} + E_{k+q})}{\omega - \epsilon_{2k} - E_k + \epsilon_{2k+q} + E_{k+q} + i \delta}} 
+ {\frac{n_F(\epsilon_{2k} - E_k) - n_F(\epsilon_{2k+q} - E_{k+q})}{\omega - \epsilon_{2k} + E_k + \epsilon_{2k+q} - E_{k+q} + i \delta}} \right], 
\label{Eq-tDDW-T-diag}
\end{eqnarray}
\begin{eqnarray}
\chi_{\text{off}}^{+-}(q,\omega)
=&{\frac {-i}{2N}} \sum_k (\frac{\epsilon_{1k} W_{k+q} + \epsilon_{1k+q} W_k }{E_k E_{k+q}}) 
\left[ -\frac{n_F(\epsilon_{2k} + E_k) - n_F(\epsilon_{2k+q} - E_{k+q})}{\omega - \epsilon_{2k} - E_k + \epsilon_{2k+q} - E_{k+q} + i \delta} 
- \frac{n_F(\epsilon_{2k} - E_k) - n_F(\epsilon_{2k+q} + E_{k+q})}{\omega - \epsilon_{2k} + E_k + \epsilon_{2k+q} + E_{k+q} + i \delta} \right. \nonumber\\
& \left. + \frac{n_F(\epsilon_{2k} + E_k) - n_F(\epsilon_{2k+q} + E_{k+q})}{\omega - \epsilon_{2k} - E_k + \epsilon_{2k+q} + E_{k+q} + i \delta}
+ \frac{n_F(\epsilon_{2k} - E_k) - n_F(\epsilon_{2k+q} - E_{k+q})}{\omega - \epsilon_{2k} + E_k + \epsilon_{2k+q} - E_{k+q} + i \delta} \right].  \label{Eq-tDDW-T-off}
\end{eqnarray}
\end{widetext}

The RPA susceptibility of the tDDW order will be
\begin{align}
\chi^{zz}_{RPA}(q,q',\omega) &= \frac{\chi_0^{zz}(q,q',\omega)} 
{ 1 - U \chi_0^{zz}(q,q',\omega)} \label{Eq-tDDWRPA1} \\
\hat{\chi}^{+-}_{RPA}(q,q',\omega) &= \sum_{q_1} \frac{\hat{\chi}^{+-}_0(q,q_1,\omega)}
{\hat{I} - U \hat{\chi}^{+-}_0(q_1,q',\omega)}, \label{Eq-tDDWRPA2}
\end{align}
where $\chi_0^{zz}(q,q',\omega)$ is from Eq.~(\ref{Eq-tDDWlong}) and $\hat{\chi}^{+-}_{0}(q,q',\omega)$ is a two by two matrix from Eq.~(\ref{Eq-tDDWtrans}).

% SDW order

For the SDW order, the Green's function matrices become
\begin{eqnarray*}
 &&{\hat G}_{\sigma}(k, i \epsilon) =\nonumber \\
& &{\frac {1}{(i \epsilon-\epsilon_{2k})^2-(E^{S}_{k})^2}}\left(
  \begin{array}{cc}
   i \epsilon +\epsilon_{1k} -\epsilon_{2k}  & \sigma \Delta_{s}          \\
    \sigma \Delta_{s}     & i \epsilon - \epsilon_{1k} -\epsilon_{2k}         \\
  \end{array}
\right).
\end{eqnarray*}

The longitudinal and transverse spin susceptibility are
\begin{align}
\chi_0^{zz}(q,q',\omega) &= \delta_{q,q'} \chi_{\text{diag}}^{zz}(q,\omega), \label{Eq-SDWlong}\\
\chi_0^{+-}(q,q',\omega) &= \delta_{q,q'} \chi_{\text{diag}}^{+-}(q,\omega) + \delta_{q,q'+Q} \chi_{\text{off}}^{+-}(q,\omega) \label{Eq-SDWtrans},
\end{align}
where $\chi^{zz}_{\text{diag}}(q,\omega)$, $\chi^{+-}_{\text{diag}}(q,\omega)$, and $\chi^{+-}_{\text{off}}(q,\omega)$ now become

\begin{widetext}
\begin{eqnarray}
\chi^{zz}_{\text{diag}}(q,\omega)
=&{\frac {1}{N}} \sum_{k} (1- {\frac{\epsilon_{1k} \epsilon_{1k+q} + \Delta_{s}^{2}}{E^{S}_k E^{S}_{k+q}}}) \left[{\frac{n_F(\epsilon_{2k} + E^{S}_k) - n_F(\epsilon_{2k+q} - E^{S}_{k+q})}{\omega - \epsilon_{2k} - E^{S}_k + \epsilon_{2k+q} - E^{S}_{k+q} + i \delta}} 
+ {\frac{n_F(\epsilon_{2k} - E^{S}_k) - n_F(\epsilon_{2k+q} + E^{S}_{k+q})}{\omega - \epsilon_{2k} + E^{S}_k + \epsilon_{2k+q} + E^{S}_{k+q} + i \delta}} \right] \nonumber \\
&+{\frac {1}{N}} \sum_k (1+ {\frac{\epsilon_{1k} \epsilon_{1k+q} + \Delta_{s}^{2} }{E^{S}_k E^{S}_{k+q}}}) \left[{\frac{n_F(\epsilon_{2k} + E^{S}_k) - n_F(\epsilon_{2k+q} + E^{S}_{k+q})}{\omega - \epsilon_{2k} - E^{S}_k + \epsilon_{2k+q} + E^{S}_{k+q} + i \delta}} 
+ {\frac{n_F(\epsilon_{2k} - E^{S}_k) - n_F(\epsilon_{2k+q} - E^{S}_{k+q})}{\omega - \epsilon_{2k} + E^{S}_k + \epsilon_{2k+q} - E^{S}_{k+q} + i \delta}} \right],  \label{Eq-SDW-L-diag} \\
\chi^{+-}_{\text{diag}}(q,\omega)
=&{\frac {1}{2N}} \sum_{k} (1- {\frac{\epsilon_{1k} \epsilon_{1k+q} - \Delta_{s}^{2}}{E^{S}_k E^{S}_{k+q}}}) \left[{\frac{n_F(\epsilon_{2k} + E^{S}_k) - n_F(\epsilon_{2k+q} - E^{S}_{k+q})}{\omega - \epsilon_{2k} - E^{S}_k + \epsilon_{2k+q} - E^{S}_{k+q} + i \delta}} 
+ {\frac{n_F(\epsilon_{2k} - E^{S}_k) - n_F(\epsilon_{2k+q} + E^{S}_{k+q})}{\omega - \epsilon_{2k} + E^{S}_k + \epsilon_{2k+q} + E^{S}_{k+q} + i \delta}} \right] \nonumber \\
&+{\frac {1}{2N}} \sum_k (1+ {\frac{\epsilon_{1k} \epsilon_{1k+q} - \Delta_{s}^{2} }{E^{S}_k E^{S}_{k+q}}}) \left[{\frac{n_F(\epsilon_{2k} + E^{S}_k) - n_F(\epsilon_{2k+q} + E^{S}_{k+q})}{\omega - \epsilon_{2k} - E^{S}_k + \epsilon_{2k+q} + E^{S}_{k+q} + i \delta}} 
+ {\frac{n_F(\epsilon_{2k} - E^{S}_k) - n_F(\epsilon_{2k+q} - E^{S}_{k+q})}{\omega - \epsilon_{2k} + E^{S}_k + \epsilon_{2k+q} - E^{S}_{k+q} + i \delta}} \right], \label{Eq-SDW-T-diag} \\
\chi_{\text{off}}^{+-}(q,\omega)
=&{\frac {\Delta_{s}}{2N}} \sum_k  
\left[ (\frac{-E^{S}_k + E^{S}_{k+q} }{E^{S}_k E^{S}_{k+q}})
 \frac{n_F(\epsilon_{2k} + E^{S}_k) - n_F(\epsilon_{2k+q} + E^{S}_{k+q})}{\omega + \epsilon_{2k} + E^{S}_k - \epsilon_{2k+q} - E^{S}_{k+q} + i \delta} 
+ (\frac{E^{S}_k - E^{S}_{k+q} }{E^{S}_k E^{S}_{k+q}}) \frac{n_F(\epsilon_{2k} - E^{S}_k) - n_F(\epsilon_{2k+q} - E^{S}_{k+q})}{\omega + \epsilon_{2k} - E^{S}_k - \epsilon_{2k+q} + E^{S}_{k+q} + i \delta}  \right. \nonumber \\
& \left. + (\frac{E^{S}_k + E^{S}_{k+q} }{E^{S}_k E^{S}_{k+q}}) 
\frac{n_F(\epsilon_{2k} + E^{S}_k) - n_F(\epsilon_{2k+q} - E^{S}_{k+q})}{\omega + \epsilon_{2k} + E^{S}_k - \epsilon_{2k+q} + E^{S}_{k+q} + i \delta} 
- (\frac{E^{S}_k + E^{S}_{k+q} }{E^{S}_k E^{S}_{k+q}})
 \frac{n_F(\epsilon_{2k} - E^{S}_k) - n_F(\epsilon_{2k+q} + E^{S}_{k+q})}{\omega + \epsilon_{2k} - E^{S}_k - \epsilon_{2k+q} - E^{S}_{k+q} + i \delta} \right].
\label{Eq-SDW-T-off}
\end{eqnarray}
\end{widetext}
The RPA susceptibility of the SDW order is in the same form as the tDDW order in Eq.~(\ref{Eq-tDDWRPA1}) and Eq.~(\ref{Eq-tDDWRPA2}).

%\bibliographystyle{apsrev4-1}
%\bibliography{susceptibility}

\end{document}